\documentclass{aa}  

\usepackage[dvipsnames]{xcolor}
\usepackage{float}  
\usepackage{graphicx}
\usepackage{txfonts}
\usepackage[]{hyperref}

\usepackage{tabularx}

\newcommand{\Msun}{\,M$_\odot$}

\newcommand\orc[1]{\href{https://orcid.org/#1}{\includegraphics[width=3mm]{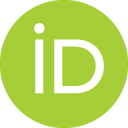}}}

\begin{document}

   \title{Twists in the flow: Revisiting convective mixing \\ in rotating stellar models}

   \subtitle{I. Effect on the stellar structure}

   \author{Poojan Agrawal \inst{1}\orc{0000-0002-1135-984X}
          \and
          Aaron Dotter \inst{2} \orc{0000-0002-4442-5700}
          \and
          Conny Aerts\inst{1,3,4}\orc{0000-0003-1822-7126}
          \and
          Le\"{i}la Bessila \inst{5} \orc{0009-0007-8721-7657}
          \and 
          St\'{e}phane Mathis\inst{5} \orc{0000-0001-9491-8012}
          }

    \institute{Institute of Astronomy, KU Leuven, Celestijnenlaan 200D, B-3001 Leuven, Belgium \label{KUL} 
    \and
    Department of Physics and Astronomy, Dartmouth College, 6127 Wilder Laboratory, Hanover, NH 03755, USA
    \and 
    Department of Astrophysics, IMAPP, Radboud University Nijmegen,
    PO Box 9010, 6500 GL Nijmegen, The Netherlands\label{Radboud}
    \and 
    Max Planck Institute for Astronomy, K\"onigstuhl 17, 69117 Heidelberg, Germany\label{MPIA}
    \and
    Université Paris-Saclay, Université Paris Cité, CEA, CNRS, AIM, Gif-sur-Yvette, F-91191, France}

   \date{}
 
  \abstract
    {Convection and rotation are both key processes in stellar evolution modelling. While standard mixing-length theory (MLT) provides a widely used modelling of convection, it neglects the effects of rotation on convective transport. }
    {We investigate how rotating mixing-length theory (R-MLT), which accounts for the influence of rotation on convection, affects the internal structure, convective mixing, and angular momentum transport in stellar models in comparison to the standard non-rotating MLT. }
    {Using the MESA stellar structure and evolution software, we model the main-sequence evolution of a 5\Msun{} star, for three cases: non-rotating, rotating with standard MLT for modelling convection, and rotating with R-MLT in convection zones, with the initial rotation rate set to 20 percent of the critical (Keplerian) value at the surface for the rotating models.}    
    {We find that R-MLT reduces both the convective velocity and mixing length in the stellar core, leading to a smaller convective diffusion coefficient and a $\sim$20 percent reduction in the extent of the convective overshooting region. While the overall size of the convective core remains nearly unchanged, R-MLT changes the resulting chemical gradient at the core–envelope boundary, shifting the peak of the Brunt–V\"ais\"al\"a frequency and modifying the angular momentum transport in that region.}
    {Including the effects of rotation in the treatment of convection through R-MLT introduces measurable structural and transport differences, underscoring the importance of incorporating rotation–convection coupling in models of stars.}


   \maketitle

\section{Introduction}
 
Convection is one of the primary mechanisms by which stars transport energy from their interior to their surface. 
It also drives chemical mixing and leaves characteristic imprints on stellar oscillation spectra \citep{Houdek2015}.
In stellar evolution models, the transport of energy and chemical elements by convection is typically described using mixing-length theory \citep[MLT;][]{Vitense1953, BohmVitense1958}. 
In MLT, energy is carried by buoyant fluid elements that rise through the stellar interior over a characteristic distance known as the mixing length, after which they dissolve into their surroundings. 
The theory assumes hydrostatic equilibrium and is solely dependent on local conditions \citep{Renzini1987}. 
Non-local and time-dependent effects, such as convective overshoot and semiconvection, are included through separate prescriptions with ad hoc free parameters
\citep[e.g.][]{Kippenhahn2012}.

Rotation is another key factor influencing stellar structure and evolution. 
It redistributes angular momentum, modifies energy transport, and affects the chemical mixture and surface elemental abundances \citep{Maeder2009}. 
In one-dimensional stellar evolution codes, rotation is commonly treated using the shellular approximation \citep{Zahn1992}, which assumes that the rotation frequency is constant on isobaric surfaces. 
Centrifugal effects are incorporated by modifying the stellar structure equations so that a spherical model of equivalent radius reproduces the volume of the oblate spheroid \citep{EndalSofia1976, Meynet1997}. 
The associated transport of angular momentum and the chemical mixing induced by rotation is then modelled either as a purely diffusive process \citep{Heger2000} or through a combined diffusion–advection formalism \citep{Zahn1992, Maeder1998, Mathis2004}.

When coupled with convection, rotation reduces convective efficiency through the Coriolis force. 
It can alter convective flow patterns \citep{Tayler1973,Stevenson1979}, modify heat transport \citep{Julien1998, Julien2012}, and influence the excitation of oscillation modes \citep{Belkacem2009, Mathis2014, Bessila2025a}. 
Although the effect of rotation on convection has been known since the pioneering work of \citet{Chandrasekhar1961}, a complete prescription parametrising the impact of rotation on stellar convection and its related transport properties over evolutionary time has not been formulated until recently.
Building on the theoretical framework developed by \citet{Stevenson1979}, \citet{Augustson2019} derived analytical prescriptions describing how key MLT variables, namely the convective velocity, mixing length, and degree of superadiabaticity (i.e. the difference between the actual and adiabatic temperature gradients), are modified in the presence of rotation. 
The approach assumes that the dominant convective mode is the one that maximises total heat flux \citep{Malkus1954}. 

The \citet{Augustson2019} formalism has since been applied, for example, to investigate the impact of rotation on the structure and evolution of fully convective low-mass stars \citep{IrelandBrowning2018}, the excitation of acoustic modes in the convective envelopes of Sun-like rotating stars \citep{Bessila2025a}, and the stability of convective layers in models of Jupiter and other gas giants \citep{Fuentes2022}.

In this work, we adopt the rotating mixing-length theory (R-MLT) formulation of \citet{Augustson2019} and \citet{Bessila2025b} to investigate the effect of rotation on convective mixing within the convective cores of intermediate-mass stars, a regime not previously explored.
We make use of Modules for Experiments in Stellar Astrophysics \citep[MESA;][]{Paxton2011,Paxton2013,Paxton2015,Paxton2018,Paxton2019,Jermyn2023}, an open-source, one-dimensional stellar structure and evolution code widely used for modelling a broad range of stellar and binary phenomena. 

Convective energy transport in MESA is modelled using MLT, with several formulations implemented, including those of \citet{BohmVitense1958}, \citet{Henyey1965}, and \citet{CoxGiuli1968}. 
More recently, time-dependent formulations of convection \citep{Kuhfuss1986, Smolec2008} have been introduced to capture the dynamic evolution of convective regions.
Rotation is incorporated in MESA using the shellular approximation \citep{Paxton2013, Paxton2019} while the transport of angular momentum and the mixing of chemical elements associated with hydrodynamical instabilities arising from differential rotation of the star are implemented in a diffusion approximation following \citet{Heger2000}. 

We used MESA to investigate how stellar rotation influences stellar structure in two scenarios: (i) when rotation and convection are treated independently and (ii) when the effects of rotation on convection are explicitly included through the R-MLT prescription.
{We deliberately restrict the scope of this work to a controlled comparison of convection prescriptions in one-dimensional stellar evolution models for a given stellar mass and do not attempt a comprehensive treatment of rotational transport, internal magnetism, or a broad grid of stellar masses. The impact of the R-MLT formulation on the broader stellar parameter space will be addressed in a dedicated follow-up study.}

The structure of this paper is as follows. 
In Sect.~\ref{sec:methods} we describe our implementation of the R-MLT prescription in MESA and outline the stellar models used in this study. 
In Sect.~\ref{sec:results} we compare the structure of rotating models with R-MLT to rotating and non-rotating models computed using standard MLT. 
In Sect.~\ref{sec:discussion} we discuss the physical implications of R-MLT in the context of stellar structure and conclude our findings in Sect.~\ref{sec:conclusions}.

\section{Methodology}
\label{sec:methods}

We used MESA version 24.08.1 to compute three sets of stellar models for a 5\,\Msun\ star: 

\begin{enumerate}
    \item 
    without any rotation and adopting standard MLT in convective regions\textbf{ (standard)}; 
    \item
    with rotation, including rotational mixing in the radiative regions and standard MLT in convective regions \textbf{(standard+rot)};
    \item 
    with rotation, using rotational mixing in the radiative regions and R-MLT in the convective zones \textbf{(standard+rot+R-MLT)}.
\end{enumerate}

\begin{figure}
\centering
    \includegraphics[width=\columnwidth,]{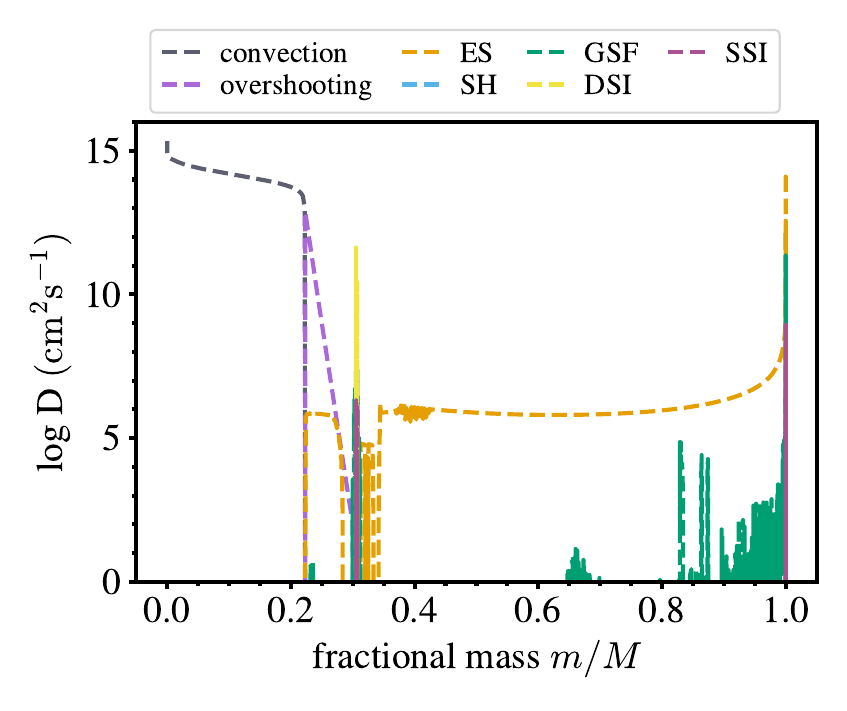} 
    \caption{Mixing coefficients as a function of enclosed mass fraction for a 5\Msun{} main-sequence star at solar metallicity, with an initial rotation rate set to 20 percent of the critical surface rotation rate. Shown are contributions from convection, convective overshoot, and rotationally- induced hydrodynamical instabilities, namely: Eddington–Sweet circulation (ES), Solberg–H\o iland instability (SH), Goldreich–Schubert–Fricke instability (GSF), dynamical shear instability (DSI), and secular shear instability (SSI).}
    \label{fig:Dmix_all}
\end{figure}

\begin{figure*}
\centering
    \includegraphics[width=\columnwidth, trim=0cm 24.6cm 0cm 0cm, clip]{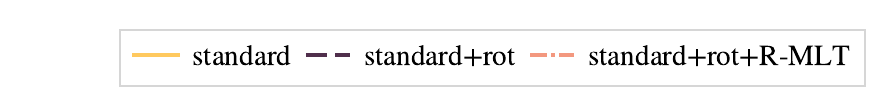}     
    \begin{tabular}{cc}
    \includegraphics[width=\columnwidth,trim=0.1cm 0.1cm 0.1cm 1.61cm, clip]{plots/MLT_vars_Xc_700.pdf} 
    & 
    \includegraphics[width=\columnwidth,trim=0.1cm 0.1cm 0.1cm 1.61cm, clip]{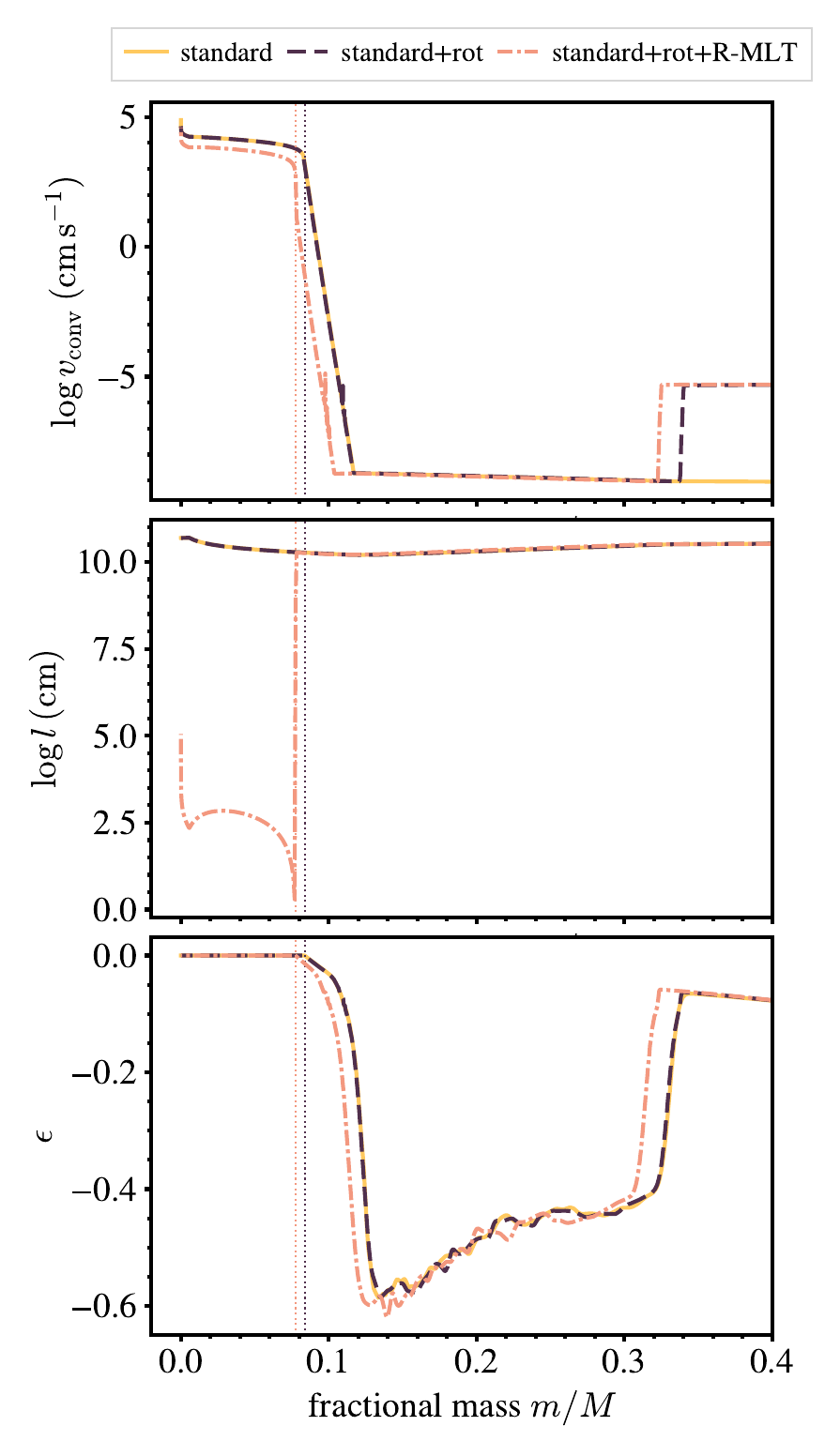}
    \end{tabular}
    \caption{ZAMS (left) and TAMS (right) profiles of the region containing the inner 40 percent in mass of the three sets of 5\Msun{} stellar models described in Sect.~\ref{sec:methods}, showing the variation of the MLT variables: convective velocity on the top, mixing length in the middle, and degree of superadiabaticity at the bottom. The solid yellow lines correspond to the non-rotating model, the dashed purple lines correspond to the model with rotation and standard MLT, and the dashed-dotted orange lines correspond to the model with rotation and R-MLT. The vertical dotted line of the corresponding colour in each panel denotes the boundary of the convective core.}
    \label{fig:mlt_vars}
\end{figure*}

\begin{figure*}
 \centering
    \includegraphics[width=\columnwidth, trim=0cm 25.2cm 0cm 0cm, clip]{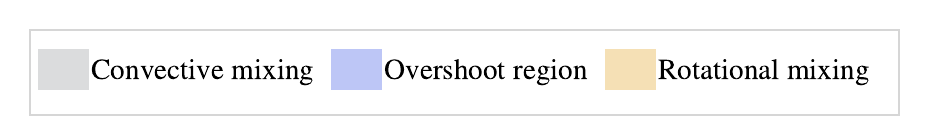} 
    \begin{tabular}{cc}
    \includegraphics[width=\columnwidth, trim=0.1cm 0.1cm 0.1cm 2.5cm, clip]{plots/Mix_profile_Xc700.pdf} 
    & 
    \includegraphics[width=\columnwidth,trim=0.1cm 0.1cm 0.1cm 2.5cm, clip]{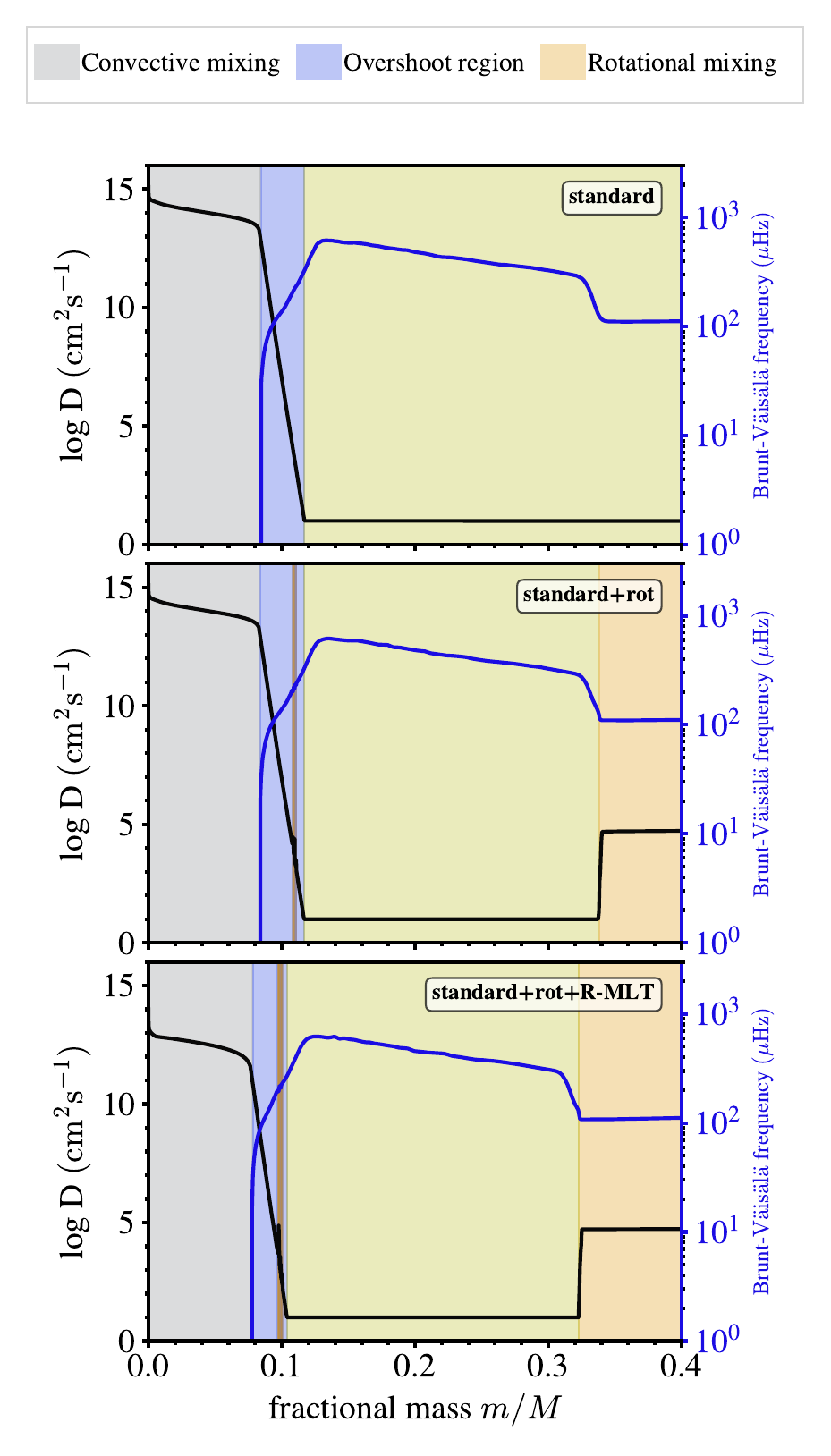}
    \end{tabular}     
    \caption{Mixing {coefficient} (black) and Brunt–V\"ais\"al\"a frequency (blue) for the same region and stellar models as shown in Fig. \ref{fig:mlt_vars}. Profiles near the ZAMS are on the left-hand side, with the non-rotating model on the top, the model with rotation and standard MLT in the middle, and the model with rotation and R-MLT at the bottom. The corresponding profiles at the TAMS are displayed in the right-hand panels. Background shading indicates the dominant mixing mechanism in different stellar layers: grey for convection, purple for overshoot, green for the radiative envelope, and orange for the envelope with rotational mixing.}
    \label{fig:mix_regions}
\end{figure*}

\begin{figure*}
\centering
    \includegraphics[width=\textwidth,trim=0cm 0cm 0cm 0cm, clip]{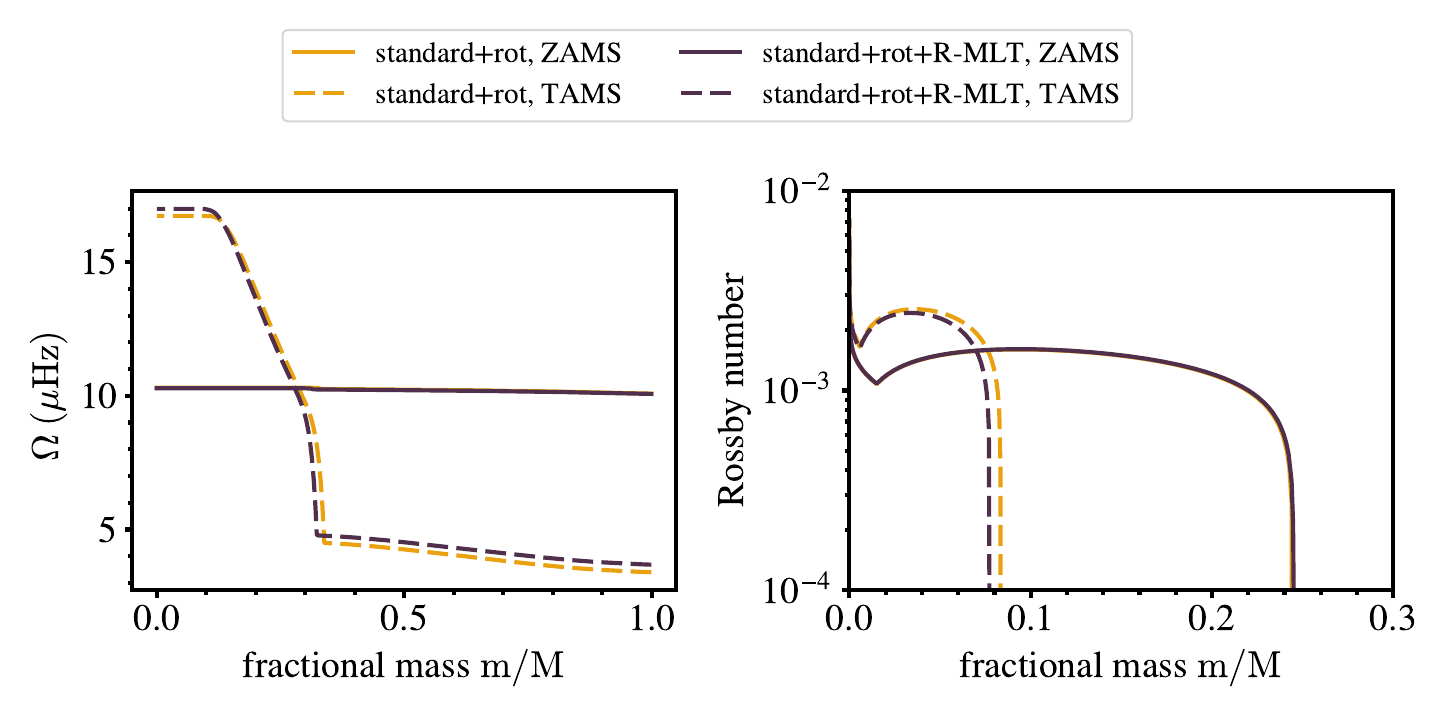} 
    \caption{Rotation frequency profiles for rotating models computed with standard MLT (dark purple) and with R-MLT (yellow) in the convective regions. The corresponding Rossby number profiles are shown in the {right} panel. Both models begin with nearly identical rotation frequency distributions near the ZAMS (solid lines) but diverge by the TAMS (dashed lines), particularly beyond the fractional mass of about 0.32.}
    \label{fig:rot_prof}
\end{figure*}
All three sets of models were computed starting from the pre-main-sequence phase with a uniform initial chemical composition from \cite{Asplund2009} at solar metallicity $\rm (Z=Z_{\odot}=0.014)$. Following \citet{Verma2019}, the initial hydrogen and helium mass fractions were set to $0.724822$ and $0.261121$, respectively. The models were evolved until the end of the main sequence.

In all three sets of models, the convective boundaries were determined from the Ledoux criterion, while the transport of convective energy was modelled using the MLT formulation by \citet{CoxGiuli1968} with a mixing-length parameter of $\alpha_{\mathrm{MLT}} = 1.8$.
Convective boundary mixing was added using an exponentially decaying overshoot prescription \citep{Freytag1996}. 
Overshoot mixing was applied until the diffusion coefficient fell below $\rm 0.05\,cm^2s^{-1}$, with overshoot parameters $f_{\rm ov}$\,=\,0.020 at convective core boundaries and $f_{\rm ov}$\,=\,0.017 at convective envelope boundaries.
Given that we only computed main-sequence models, we ignored semiconvection and thermohaline instabilities. We enforced a minimum level of envelope mixing of $\rm 10\,cm^2s^{-1}$ in the radiative regions.

For both sets of rotating models, once the nuclear-burning luminosity reached 90 percent of the surface luminosity during the final contraction from the pre-main sequence to the zero-age main sequence (ZAMS), the rotation rate was gradually increased in 50 time steps until it reached 20 percent of the surface critical (Keplerian) value. 
During this spin-up phase, the time step control was tightened (\texttt{time\_delta\_coeff}= 0.1, compared to 0.5 for the rest of the evolution) to ensure that the ZAMS structure and subsequent evolution were insensitive to the ramp duration \footnote{{We find no difference in the structure or the evolution of the star by altering the ramp duration to 20 steps instead of 50.}}.

The Eddington–Sweet meridional circulation was included as the sole mechanism for angular-momentum transport and chemical mixing due to rotation. Other hydrodynamical instabilities were excluded, as they introduced numerical noise in rotational and chemical profiles while negligibly contributing to overall transport processes in our models {(Fig~\ref{fig:Dmix_all})}. Rotational mixing was applied only in radiative regions and not within convective zones.

The rotating mixing-length theory was implemented following the prescription of \citet{Augustson2019}. We provide a summary of our implementation, but refer interested readers to \citet{Augustson2019} for further details and derivations (see also the appendix of \citealt{Bessila2025a}).
{To include R-MLT, the MLT parameters, namely convective velocity $v_{\rm conv,0}$, mixing length $l_0$, and the difference between the actual and adiabatic temperature gradient, called superadiabaticity $(\epsilon_0=\nabla-\nabla_{\rm ad})$, were initially calculated using the \citet{CoxGiuli1968} formulation of the standard MLT prescription.}

For every mass shell in the star where $v_{\rm conv,0}$, exceeds $10^{-12}\,\mathrm{cm\,s^{-1}}$ and the rotation frequency $\Omega_0$ exceeds $10^{-12}\,\mathrm{rad\,s^{-1}}$ -- that is, where both quantities are non-negligible -- we solved the following equation for the normalised wavevector $z$
\begin{equation}
2 z^5-5 z^2-\frac{18 \cos ^2 \theta}{25 \pi^2 \mathrm{Ro}^2 \tilde{s}^2}=0 , 
\label{eq:R-MLT}
\end{equation}
{with $\tilde{s}=2^{1 / 3}\cdot 3^{1 / 2} \cdot 5^{-5 / 6}$}.

$\theta$ denotes the angle between local effective gravity (the vertical direction) and the rotation vector. Following \citet{Bessila2025a}, we assume that the rotation axis is aligned with the polar axis, therefore $\cos\theta$=1.
The dimensionless parameter Ro is the convective Rossby number, which measures {the ratio of the advection of convective flows to their}
Coriolis acceleration. It is defined as
\begin{equation}
\mathrm{Ro} = \frac{v_{\rm conv,0}}{2 \Omega_0 l_0},
\label{eq:rossby}
\end{equation}
where $v_{\rm conv,0}$ and $l_0$ are the convective velocity and mixing length computed from the unmodified (i.e. non-rotationally corrected) MLT expressions, and $\Omega_0$ is the local angular velocity from the rotating stellar model.
 
Equation~\ref{eq:R-MLT} has only one real root, $z_0$. This root is determined numerically using the Newton–Raphson method. Once $z_0$ is known, the convective velocity in the presence of rotation can be calculated using the following relation
\begin{equation}
v_{\rm conv,\Omega}=v_{\rm conv,0} \cdot 5 \frac{\tilde{s}}{\sqrt{6 z_0}}.
\label{eq:conv_v}
\end{equation}
Similarly, the mixing length taking into account the rotation of the star is given by
\begin{equation}
l_{\Omega}=l_0 \cdot {\sqrt{\frac{5}{2 z_0^{3}}}}.
\label{eq:l}
\end{equation}
Finally, superadiabaticity is calculated from its original value in the absence of rotation, $\epsilon_0$, using the following equation:
\begin{equation}
\epsilon_{\Omega}={\epsilon_0} \cdot \frac{25 \pi^2 \mathrm{Ro}^2 \tilde{s}^2 z_0^5+6 }{25 \pi^2 \mathrm{Ro}^2\left(z_0^3-1\right)}.
\label{eq:superadiabtaicity}
\end{equation}
The scaled R-MLT values are calculated for every mass shell in the star whenever the convective transport of energy is evaluated, accounting for the effect of rotation on the convective transport of energy and hence on the evolution of the star. 

For fast rotators (${\rm Ro\ll 1}$), Eq. \ref{eq:R-MLT} reduces to ${\rm z \propto Ro^{-2/5}}$, yielding ${\rm v_{\rm conv,\Omega}/v_{\rm conv,0} \propto Ro^{1/5} , l_{\Omega}/l_0 \propto Ro^{3/5}}$, and ${\rm \epsilon_{\Omega}/{\epsilon_0} \propto Ro^{-4/5}}$. 
These scalings indicate that convection becomes slower, smaller-scale, and less efficient. Conversely, for slow rotators (${\rm Ro\gg 1}$), these quantities asymptotically approach unity, and convection is effectively insensitive to rotation.

\section{Results}
\label{sec:results}

In this section we examine the structural properties of our 5\Msun{} stellar models and how these are affected by the use of R-MLT in the convective regions.
Figure~\ref{fig:mlt_vars} shows the convective velocity, mixing length, and superadiabaticity within the inner 40 percent of the stellar mass for all three sets of models described in Sect.~\ref{sec:methods}, both near the (ZAMS, $X_c = 0.7$) and the terminal-age main sequence (TAMS, $X_c = 10^{-3}$).
At the ZAMS, the convective core accounts for almost 25 percent of the stellar mass in all three sets of models. The standard model exhibits large convective velocities and mixing lengths within the convective core. 
Outside the core, the convective velocity declines exponentially due to overshooting, eventually reaching a minimum imposed by the minimum mixing value enforced in the envelope. 

\begin{table*}
\centering
\caption{Change in overshooting-region mass fraction in rotating models. }
\label{tab:model_var}
\begin{tabular}{|c|c|c|c|c|c|}
\hline
Latitudinal & Minimum radiative-zone  & Exponential overshoot & Rotational  & $\Delta M_{\rm ov}/M$ & $\Delta M_{\rm ov}/M$  \\
dependence ($\theta$) & mixing  (\texttt{min\_D\_mix}) &  threshold (\texttt{overshoot\_D\_min}) & mixing  & (ZAMS, \%) & (TAMS, \%) \\
\hline
$\cos\theta$\,=\,1 & 10 & 0.05 & ES & 18.7 & 21.6 \\
\hline
$\boldsymbol{\langle \cos^2\theta \rangle = \frac{1}{3}}$ & 10 & 0.05 & ES & 16.6 & 19.6 \\
$\cos\theta$\,=\,1 & \textbf{0}& 0.05 &  ES & 20.6 & 19.5 \\
$\cos\theta$\,=\,1 & \textbf{0} & \textbf{1} & ES & 20.6 & 19 \\
$\cos\theta$\,=\,1 & 10 & 0.05 & \textbf{ES+SH+GSF+} & 18.7 & 21.6 \\
                    &     &     & \textbf{DSI+SSI} &    &   \\
\hline
\end{tabular}
 \tablefoot{$\Delta M_{\rm ov}/M$, is the relative change in overshooting-region mass fraction due to R-MLT under different model assumptions. The first row corresponds to the fiducial model; deviations in subsequent rows are highlighted in bold.}
\end{table*}

The standard+rot model, which incorporates rotation without any modifications to MLT, introduces no significant structural changes, apart from a modest increase in the convective velocity in the envelope due to the rotationally induced mixing.  
In contrast, application of the R-MLT formulation (standard+rot+R-MLT model) leads to a marked reduction in both the convective velocity and the mixing length throughout the convective core region.
Although the deviations in mixing length vanish at the convective boundary, the reduction in convective velocity extends into the overshooting region.

As the models evolve, the convective core recedes, as illustrated in the right-hand panels of Fig.~\ref{fig:mlt_vars}, showing the results of the model at the TAMS. 
The general trends observed for the ZAMS remain, but the effects of the R-MLT formulation become more pronounced. 
In particular, differences in mixing length and velocity occur in the region left behind by the retreating convective core.
Although these differences are significant, they do not affect the overall behaviour of the convective core, as illustrated in Fig.~\ref{fig:mix_regions}, which shows the mixing profile and Brunt–V\"ais\"al\"a frequency for the same mass region as in Fig.~\ref{fig:mlt_vars}. 
This is because the superadiabaticity is close to zero throughout the convective and overshooting regions (bottom panels in Fig.~\ref{fig:mlt_vars}), indicating that convection remains highly efficient and the convective boundaries are not substantially altered.

However, the same does not hold for convective boundary mixing. Since the mixing coefficient, $D_{\rm conv}$, is proportional to the product of convective velocity and mixing length, its reduction in the R-MLT models results in smaller mixing coefficients throughout the convective and overshooting zones.
Consequently, the extent of the overshooting region decreases to about 7 percent of the stellar mass in the R-MLT models, compared to nearly 9 percent in both standard and standard+rot models near the ZAMS.

At the TAMS (right panels of Fig.~\ref{fig:mix_regions}), the effects of R-MLT during earlier evolution begin to reveal differences in the convective region as well. 
In the R-MLT case, the convective region decreases to 7.9 percent, while it is approximately 8.3 percent in the other two cases. 
Similarly, the overshooting region in the R-MLT models shrinks to 2.6 percent of the stellar mass, compared to 3.2 percent in the non-rotating and standard MLT rotating models.

The smaller overshooting region in the R-MLT models causes the chemical gradient near the core–envelope boundary to shift inward toward the stellar core. Since the Brunt–V\"ais\"al\"a frequency, $N$, at any location within the star depends on the combined effects of superadiabaticity and the chemical gradient, its peak is also shifted inward compared to rotating models computed with standard MLT.

A similar trend is seen in the transport of angular momentum, as illustrated in Fig.~\ref{fig:rot_prof}, which shows the rotation frequency (left panel) and Rossby number (Eq.~\ref{eq:rossby}, right panel) for both sets of rotating models (standard+rot and standard+rot+R-MLT). Near the ZAMS, both models exhibit nearly solid-body rotation with similar rotation frequency profiles. As the star evolves, the rotation becomes more differential, and the differences in diffusion coefficients introduced by the R-MLT prescription become significant, leading to a divergence of the rotation frequency profile near the core–envelope boundary, at a fractional mass between 0.3 and 0.35 at the TAMS.

\section{Discussion}
\label{sec:discussion}

\begin{figure*}
\centering
    \includegraphics[width=0.95\columnwidth]{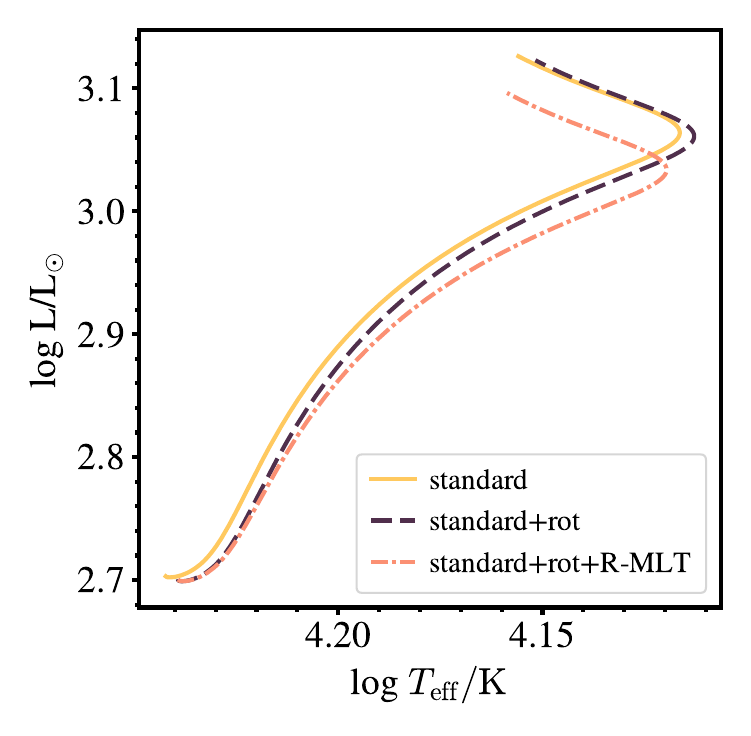} 
    \includegraphics[width=\columnwidth]{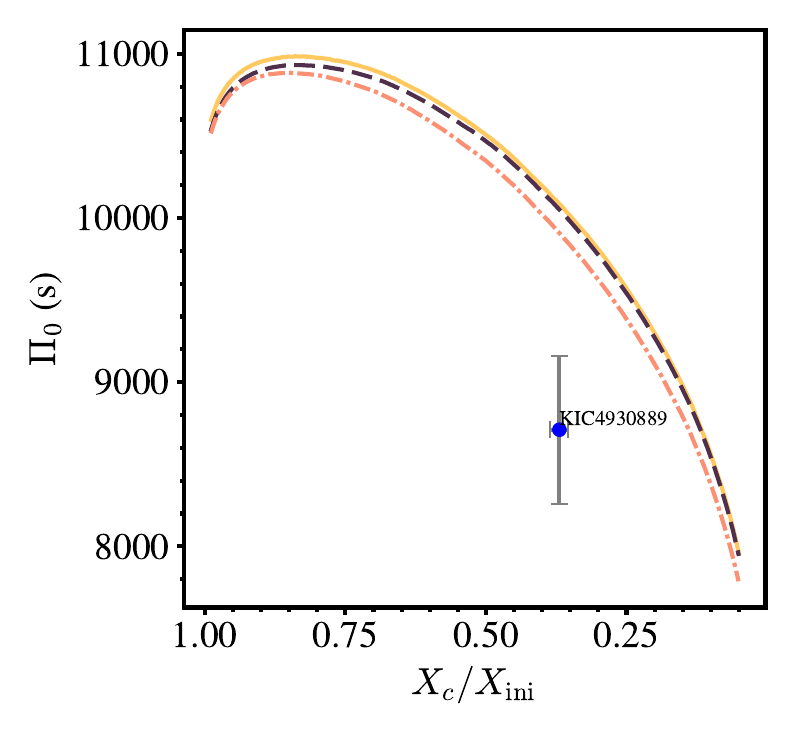} 
    
    \caption{\textbf{Left panel}: Hertzsprung-Russell (HR) diagram showing the main-sequence evolution of a 5\Msun{} star. \textbf{Right panel}: Evolution of the buoyancy travel time as a function of the hydrogen mass fraction. The solid yellow line represents the non-rotating model computed with standard MLT in the convective zones; the dashed purple line corresponds to the rotating model with rotational mixing in the radiative zones and standard MLT in the convective zones; and the dash-dotted orange line shows the rotating model but with R-MLT instead of standard MLT in the convective zones. }
    \label{fig:hrd}
\end{figure*}

\begin{table*}
\centering
\caption{Stellar parameters for models at different evolutionary stages.
}
\label{tab:rep_values}
\begin{tabular}{|l|l|l|l|l|l|l|l|}
\hline
$X_c/X_{\rm ini}$ & Model & log L\,(L$_\odot$) & R\,(R$_\odot$)  & Age (Myr) & $M_{\rm cc}$ (M$_\odot$) & $\Pi_0$ (s)\\
\hline
    & standard & 2.713 & 2.583 & 8.44 & 1.225 & 10777.6 \\
0.9 & standard+rot & 2.710 & 2.606 & 8.51 & 1.223 & 10721.0 \\
    & standard+rot+R-MLT & 2.709 & 2.606 & 8.19 & 1.225 & 10700.3 \\
\hline
     & standard & 2.896 & 3.749 & 75.10  & 0.903 & 10632.6 \\
 0.5 & standard+rot & 2.893 & 3.795 & 75.38 & 0.901 & 10588.0 \\
     & standard+rot+R-MLT & 2.881 & 3.741 & 72.29 & 0.887 & 10478.2 \\
\hline
     & standard & 3.055 & 6.518 & 110.03 & 0.525 & 8228.8 \\
 0.1 & standard+rot & 3.052 & 6.607 & 110.45 & 0.525 & 8196.9 \\
     & standard+rot+R-MLT & 3.027 & 6.239 & 106.44 & 0.509 & 8040.3 \\
\hline
\end{tabular}
\tablefoot{The listed quantities are stellar luminosity ($\log L$), stellar radius ($R$), age, convective core mass ($M_{\rm cc}$), and buoyancy travel time ($\Pi_0$). The three evolutionary stages correspond to models near the ZAMS ($X_c/X_{\rm ini}=0.9$), mid–main sequence ($X_c/X_{\rm ini}=0.5$), and near the TAMS ($X_c/X_{\rm ini}=0.1$). The evolutionary stage is indicated by the central hydrogen mass fraction $X_c/X_{\rm ini}$.}
\end{table*}

The differences among the three sets of stellar models discussed in Sect.~\ref{sec:results} can be traced primarily to variations in chemical mixing outside the convective boundary. For both sets of rotating and non-rotating models that use standard MLT in the convective zones, the main difference arises from rotational mixing via the Eddington–Sweet meridional circulation modelled in a diffusive approach. For a rotation rate of 20 percent of the critical value, rotational mixing due to Eddington–Sweet circulation has an average diffusion coefficient of $\rm 10^5\,cm^2s^{-1}$ (Fig.~\ref{fig:mix_regions}). While this is significant on its own, it is orders of magnitude weaker than mixing due to convection and convective overshoot, which have average values of $\rm 10^{15}\,cm^2s^{-1}$ and $\rm 10^{11}\,cm^2s^{-1}$, respectively. Consequently, adding rotation without any modifications to MLT produces only minor differences in chemical gradients and the overall evolution of the star, as shown in Fig.~\ref{fig:hrd}. 

For models with rotation using R-MLT in the convective region, the reduced overshooting region has a more pronounced effect on chemical mixing beyond the convective boundary. The resulting change in chemical gradients, in turn, changes the way other variables such as the Brunt–V\"ais\"al\"a and rotation frequencies evolve. Although the magnitude of the difference in variables is modest at the ZAMS, the effect accumulates throughout the main-sequence evolution, becoming more noticeable towards the TAMS (see Fig.~\ref{fig:rotmlt_zams} for relative differences in various physical variables between the rotating model computed with standard MLT and the one with R-MLT at the ZAMS, and Fig.~\ref{fig:rotmlt_tams} at the TAMS). Figures~\ref{fig:rot_zams} and \ref{fig:rot_tams} show the relative differences between rotating and non-rotating models at the ZAMS and TAMS, respectively.

Despite the differences in mixing coefficients, the overall structure of the convective core remains largely unchanged between the rotating and non-rotating MLT models.
The near-zero superadiabaticity observed throughout the convective and overshooting zones in all models confirms that convection remains highly efficient and that the main distinction introduced by R-MLT lies in modifying the convective flux balance and associated chemical gradients.
These trends are consistent with the predictions of \citet{Augustson2019}, three-dimensional simulations of rotating convection by \citet{Korre2021}, and MESA-based studies of fully convective low-mass stars (\citealt{IrelandBrowning2018}).

We tested the robustness of our results to several input-physics choices, {as listed in Table~\ref{tab:model_var}}. 
Our analysis accounts only for Eddington–Sweet circulation as a source of rotational mixing. We find that including additional hydrodynamic rotational instabilities available in MESA does not alter the main-sequence results presented here. Nonetheless, these instabilities may become relevant in later evolutionary phases, where their contribution to angular-momentum and chemical transport can be more significant \citep{Talon1997, Maeder2000}.

Convective boundary criteria can also influence the impact of R-MLT and semiconvective mixing may, in principle, modify the chemical gradient outside the retreating core. However, in our main-sequence models, the radiative temperature gradient remains {smaller} than the adiabatic gradient throughout the radiative interior, except within fully convective regions identified by the Ledoux criterion {(Fig.~\ref{fig:semiconvection})}. Enabling semiconvective mixing, or alternatively adopting the Schwarzschild criterion to determine the convective boundary, produces no change in our results. We note, however, that these effects may become relevant at later evolutionary stages or in more massive main-sequence stars. 

Our use of $\cos\theta$\,=\,1 in Eq.~\ref{eq:R-MLT} corresponds to the maximal rotational constraint in the R-MLT formulation. {To explore the effect of the latitudinal dependence, we consider the spherical average of $\cos^2\theta$ appearing in Eq.~\ref{eq:R-MLT}, i.e. $ \langle \cos^2\theta \rangle = \frac{1}{3}$. This leads to only a modest reduction in the absolute diffusion coefficient and, consequently, in the overshooting region mass fraction.}
Thus, adopting the spherical-average prescription does not significantly alter the relative differences between the rotating models computed with and without R-MLT.

We further vary the diffusion-coefficient threshold, \texttt{overshoot\_D\_min}, that sets the extent of exponential overshoot in MESA. 
To prevent it from interfering with this comparison, we also turn off minimum mixing in radiative regions (\texttt{min\_D\_mix}\,=\,0).
As expected, a lower threshold increases the absolute extent of the overshoot region; however, the relative reduction in the overshoot region induced by R-MLT remains nearly unchanged. 
These tests demonstrate that the relative reduction in overshoot produced by R-MLT is insensitive to the specific value adopted, within reasonable limits.

Our results are independent of the specific MLT formulation but depend on the adopted mixing-length parameter. Larger mixing leads to stronger suppression under the R-MLT framework, resulting in reduced convective velocities and mixing coefficients. This suppression may significantly influence chemical and angular momentum transport, with potential implications for the later evolutionary phases of intermediate- and high-mass stars that possess convective cores on the main sequence. 

From an observational perspective, the reduced overshoot mixing predicted by the R-MLT formulation could have several detectable consequences. A smaller overshooting region compared to rotating models computed with standard MLT implies a reduced fuel supply to the convective core, shortening the main-sequence lifetime, as well as lowering the luminosity and radius of the star, as shown in the left panel of Fig.~\ref{fig:hrd}, and further quantified in Table~\ref{tab:rep_values}. The impact of the smaller overshooting region becomes progressively more pronounced as the stars evolve along the main sequence, leading to increasing differences in the stellar parameters.

In addition, the shift in the the Brunt–V\"ais\"al\"a frequency profile may produce detectable differences in g-mode period spacings that are highly sensitive to internal chemical stratification, as predicted by \citet{Pedersen2018} and \citet{Michielsen2019}. 
Such features can have strong asteroseismic signatures, as has been measured in moderate to rapid rotators belonging to the pulsation classes of slowly pulsating B (SPB) stars  \citep{Papics2015,Papics2017,Szewczuk2018,Pedersen2021,Szewczuk2021} and the $\gamma$ Doradus stars \citep{VanReeth2015,GangLi2019,GangLi2020}.
This is illustrated in the right panel of Fig.~\ref{fig:hrd}, which shows the evolution of the buoyancy travel time $\Pi_0$ \citep[as defined in ][]{Aerts2021}, as a function of $X_c/X_{\mathrm{ini}}$ along the main sequence for the three sets of models. 
The differences in $\Pi_0$ between standard+rot and standard+rot+R-MLT are on the order of tens of seconds near ZAMS ($X_c/X_{\mathrm{ini}}=0.9$), increasing to hundreds of seconds by mid main sequence ($X_c/X_{\mathrm{ini}}=0.5$) to late main sequence ($X_c/X_{\mathrm{ini}}=0.1$).
For comparison, we also plot the measured values of $\Pi_0$ versus $X_c/X_{\mathrm{ini}}$ {and the corresponding 1-$\sigma$ uncertainties} for the SPB star KIC 4930889 \citep{Pedersen2021,Pedersen2022}, a 4.4\Msun{} star with one of the smallest uncertainties in $\Pi_0$ from its four-year Kepler light curve. 
{The value of $\Pi_0$ for KIC 4930889 has been calculated by fitting asymptotic period spacing patterns \citep{VanReeth2015}, and is model-independent, while $X_c/X_{\mathrm{ini}}$ has been estimated through forward asteroseismic stellar modelling of the period spacing patterns, and is dependent on the stellar evolution models.}

In this case, the structural differences induced by R‑MLT are smaller than the observational uncertainties in $\Pi_0$, and therefore are unlikely to be distinguishable within the precision of four-year Kepler observations.
A detailed asteroseismic modelling of individual mode frequencies is deferred to a dedicated follow-up study, which will investigate whether subtle structural differences induced by R‑MLT can be detected when fitting individual g-mode frequencies, rather than $\Pi_0$ as a global diagnostic, thereby providing a more sensitive probe of internal chemical stratification.

\section{Conclusions}
\label{sec:conclusions}

We investigated the impact of rotating mixing-length theory (R-MLT) from \cite{Augustson2019} on the internal structure and convective properties of a 5\Msun{} main-sequence star at solar metallicity, with an initial rotation frequency set to 20 percent of the critical surface rotation frequency. 
We modified the one-dimensional stellar evolution code MESA to implement the new R-MLT scaling prescriptions from \citet{Bessila2025a} and compared three sets of models: non-rotating models with standard MLT, rotating models with rotational mixing and standard MLT in the convective region, and rotating models employing R-MLT in the convective region.
Our main findings are as follows:
\begin{enumerate}
    \item For models using standard MLT, rotational mixing via Eddington–Sweet meridional circulation, modelled within a diffusive approach, is much smaller compared to convective and overshoot mixing, resulting in minimal differences in chemical gradients and structural properties between rotating and non-rotating models. 
    
    \item The use of R-MLT leads to systematically lower convective velocities and mixing lengths within the convective core compared to the case where standard non-rotating MLT is used. As the diffusion coefficient scales with the product of these quantities, R-MLT models exhibit overshooting regions that are reduced by approximately 20 percent, relative to standard MLT models.

    \item Despite significant local differences, the overall extent of the convective core remains similar across all models, owing to the nearly adiabatic stratification and efficient convective transport.

    \item The shift in the chemical gradient left behind by the weaker core boundary mixing modifies the local buoyancy, producing a shift in the peak of the Brunt–V\"ais\"al\"a frequency, as well as altering the evolution of the internal rotation frequency profile along the main sequence.
\end{enumerate}

Our results demonstrate that R-MLT systematically affects the efficiency of mixing in 5\Msun{} stars, highlighting the importance of incorporating such prescriptions into stellar evolution models and in high-precision asteroseismology. Future work will extend this analysis to a wider range of stellar masses and rotation rates, to assess the overall impact of R-MLT across the HR diagram, and investigate the combined effect of magnetic fields and rotation on stellar convection.

\begin{acknowledgements}
    {We thank the referee for their constructive and detailed comments. We thank Joey Mombarg, Mathijs Vanrespaille and Jelle Vandersnickt for useful comments and discussions.} The authors acknowledge support from the European Research Council (ERC) under the Horizon Europe programme (Synergy Grant agreement 101071505: 4D-STAR). While partially funded by the European Union, views and opinions expressed are, however, those of the authors only and do not necessarily reflect those of the European Union or the European Research Council. Neither the European Union nor the granting authority can be held responsible for them. S.M. and L.B. acknowledge support from the CNES PLATO grant @ DAp/CEA.
\end{acknowledgements}

\bibliographystyle{aa} 
\bibliography{bibliography} 

\begin{appendix}
\onecolumn
\section{Additional figures}
\label{sec:appendix}

\begin{figure*}[!hbtp]
    \centering
    \begin{tabular}{ccc}
     \includegraphics[width=0.32\textwidth, page=1, trim=0.65cm 0.65cm 0.65cm 0.65cm, clip]{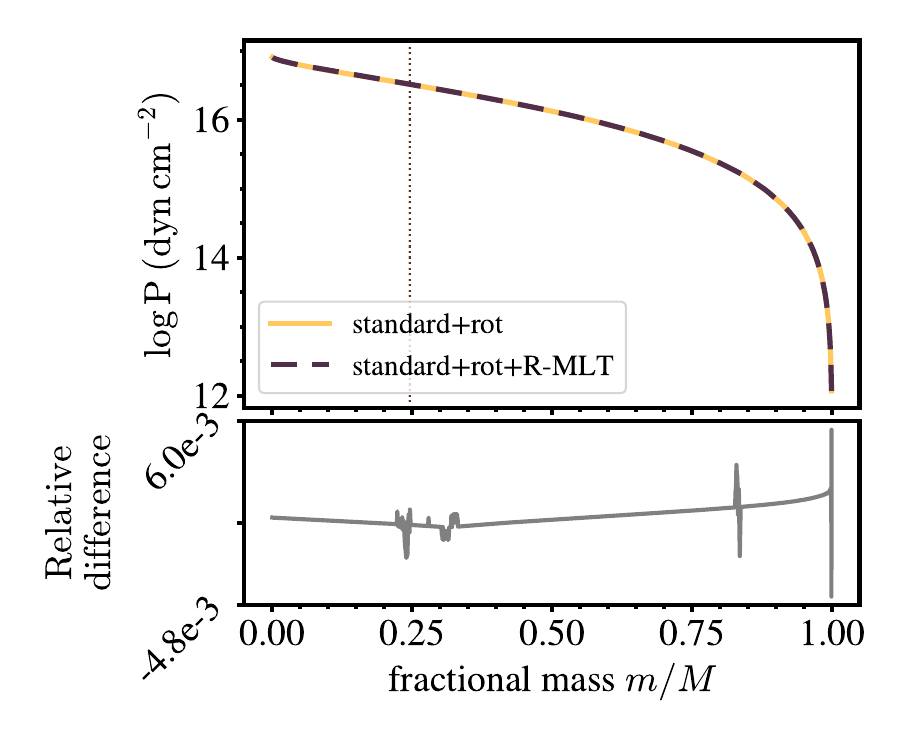}
    \includegraphics[width=0.32\textwidth, page=2, trim=0.65cm 0.65cm 0.65cm 0.65cm, clip]{plots/Residual_rotMLT_Xc_700.pdf}
    \includegraphics[width=0.32\textwidth, page=3, trim=0.65cm 0.65cm 0.65cm 0.65cm, clip]{plots/Residual_rotMLT_Xc_700.pdf}
    \\
    \includegraphics[width=0.32\textwidth, page=8, trim=0.65cm 0.65cm 0.65cm 0.65cm, clip]{plots/Residual_rotMLT_Xc_700.pdf}
    \includegraphics[width=0.32\textwidth, page=9, trim=0.65cm 0.65cm 0.65cm 0.65cm, clip]{plots/Residual_rotMLT_Xc_700.pdf}
    \includegraphics[width=0.32\textwidth, page=10, trim=0.65cm 0.65cm 0.65cm 0.65cm, clip]{plots/Residual_rotMLT_Xc_700.pdf}  
    \\
     \includegraphics[width=0.32\textwidth, page=4, trim=0.65cm 0.65cm 0.65cm 0.65cm, clip]{plots/Residual_rotMLT_Xc_700.pdf}  
    \includegraphics[width=0.32\textwidth, page=5, trim=0.65cm 0.65cm 0.65cm 0.65cm, clip]{plots/Residual_rotMLT_Xc_700.pdf}
    \includegraphics[width=0.32\textwidth, page=11, trim=0.65cm 0.65cm 0.65cm 0.65cm, clip]{plots/Residual_rotMLT_Xc_700.pdf}
    \end{tabular}
   
    \caption{Internal structure of a 5\Msun{} ZAMS model with initial rotation frequency 20 percent of the critical frequency at the surface, computed with standard MLT (solid yellow lines) and R-MLT (dashed purple lines). Each of the nine panels shows pressure ($\log P$), temperature ($\log T$), density ($\log \rho$), chemical gradient ($\nabla_{\mu}$), actual temperature gradient ($\nabla_{\mathrm{T}}$), adiabatic temperature gradient ($\nabla_{\mathrm{ad}}$), Brunt-V\"ais\"al\"a frequency ($N$), mean molecular weight per particle (ions + free electrons, $\mu$), and fractional rotation frequency ($\Omega/\Omega_{\mathrm{crit}}$) as functions of fractional mass. Upper subplots show the absolute profiles for both models, and lower subplots show their relative differences, $(X_{\mathrm{rot+R-MLT}}-X_{\mathrm{rot}})/X_{\mathrm{rot}}$. Vertical dotted lines of the corresponding colour in the upper plots show the convective boundary for each model. Surface points have been removed for clarity.}
    
    \label{fig:rotmlt_zams}
\end{figure*}

\begin{figure*}
    \centering
    \begin{tabular}{lll}
    \includegraphics[width=0.32\textwidth, page=1, trim=0.65cm 0.65cm 0.65cm 0.65cm, clip]{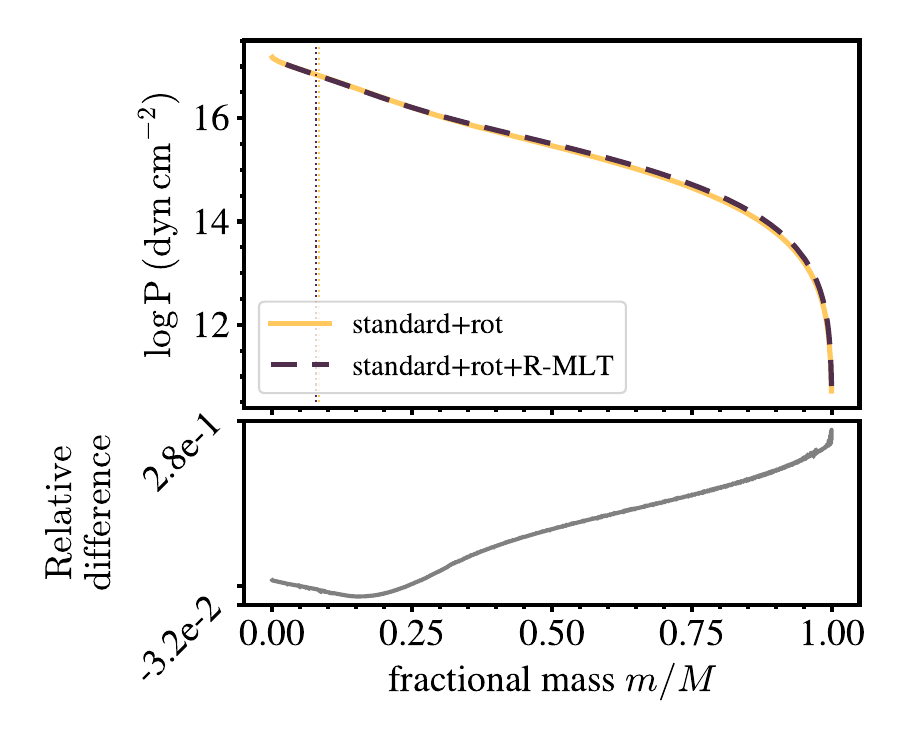} 
    \includegraphics[width=0.32\textwidth, page=2, trim=0.65cm 0.65cm 0.65cm 0.65cm, clip]{plots/Residual_rotMLT_Xc_001.pdf}
    \includegraphics[width=0.32\textwidth, page=3, trim=0.65cm 0.65cm 0.65cm 0.65cm, clip]{plots/Residual_rotMLT_Xc_001.pdf}
    \\
    \includegraphics[width=0.32\textwidth, page=8, trim=0.65cm 0.65cm 0.65cm 0.65cm, clip]{plots/Residual_rotMLT_Xc_001.pdf}
    \includegraphics[width=0.32\textwidth, page=9, trim=0.65cm 0.65cm 0.65cm 0.65cm, clip]{plots/Residual_rotMLT_Xc_001.pdf}
    \includegraphics[width=0.32\textwidth, page=10, trim=0.65cm 0.65cm 0.65cm 0.65cm, clip]{plots/Residual_rotMLT_Xc_001.pdf}  
    \\
     \includegraphics[width=0.32\textwidth, page=4, trim=0.65cm 0.65cm 0.65cm 0.65cm, clip]{plots/Residual_rotMLT_Xc_001.pdf}  
    \includegraphics[width=0.32\textwidth, page=5, trim=0.65cm 0.65cm 0.65cm 0.65cm, clip]{plots/Residual_rotMLT_Xc_001.pdf}
    \includegraphics[width=0.32\textwidth, page=11, trim=0.65cm 0.65cm 0.65cm 0.65cm, clip]{plots/Residual_rotMLT_Xc_001.pdf}
    \end{tabular}
   
    \caption{Same as Fig.~\ref{fig:rotmlt_zams} but at TAMS. As the star evolves through the main sequence, differences between rotating models using standard MLT (standard+rot) and R-MLT (standard+rot+R-MLT) become more pronounced due to the differing amounts of overshooting. The difference in overshooting alters the chemical profile near the core-envelope boundary, which in turn drives the deviations between the models.
   }
    
    \label{fig:rotmlt_tams}
\end{figure*}

\begin{figure*}
    \centering
    \begin{tabular}{lll}
     \includegraphics[width=0.32\textwidth, page=1, trim=0.65cm 0.65cm 0.65cm 0.65cm, clip]{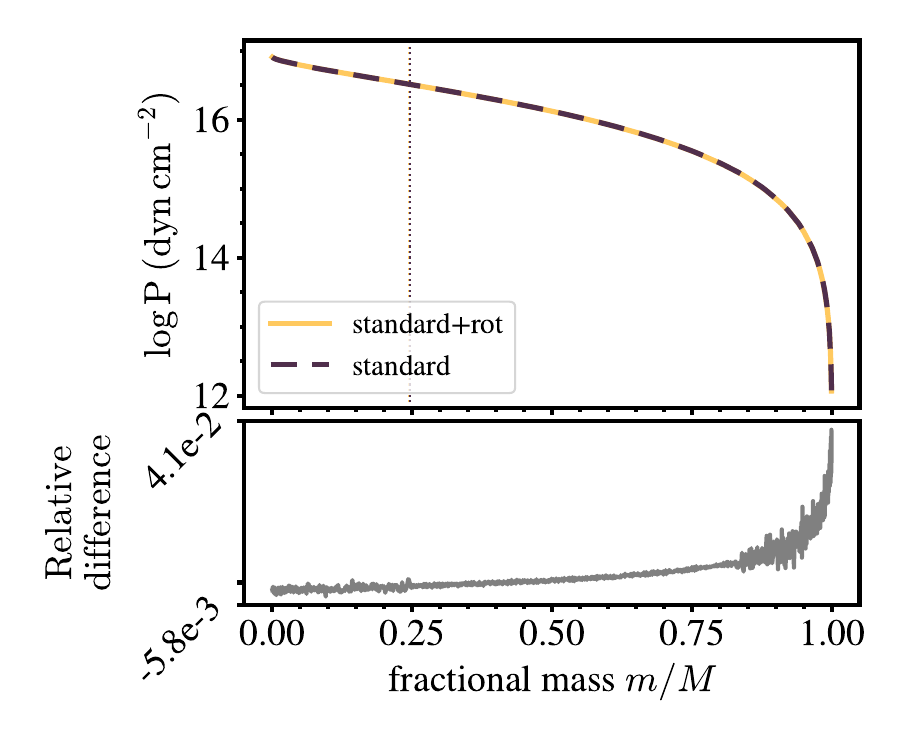}
    \includegraphics[width=0.32\textwidth, page=2, trim=0.65cm 0.65cm 0.65cm 0.65cm, clip]{plots/Residual_rot_Xc_700.pdf}
    \includegraphics[width=0.32\textwidth, page=3, trim=0.65cm 0.65cm 0.65cm 0.65cm, clip]{plots/Residual_rot_Xc_700.pdf}
    \\
    \includegraphics[width=0.32\textwidth, page=8, trim=0.65cm 0.65cm 0.65cm 0.65cm, clip]{plots/Residual_rot_Xc_700.pdf}
    \includegraphics[width=0.32\textwidth, page=9, trim=0.65cm 0.65cm 0.65cm 0.65cm, clip]{plots/Residual_rot_Xc_700.pdf}
    \includegraphics[width=0.32\textwidth, page=10, trim=0.65cm 0.65cm 0.65cm 0.65cm, clip]{plots/Residual_rot_Xc_700.pdf}  
    \\
     \includegraphics[width=0.32\textwidth, page=4, trim=0.65cm 0.65cm 0.65cm 0.65cm, clip]{plots/Residual_rot_Xc_700.pdf}  
    \includegraphics[width=0.32\textwidth, page=5, trim=0.65cm 0.65cm 0.65cm 0.65cm, clip]{plots/Residual_rot_Xc_700.pdf}
    \end{tabular}
   
    \caption{Same as Fig.~\ref{fig:rotmlt_zams} but comparing models computed with rotation (standard+rot; solid yellow) and without rotation (standard; dashed purple) at ZAMS. With an initial rotation frequency of twenty percent of the surface critical rotation frequency, most parameters differ by less than one percent between the rotating and non-rotating models. }
    
    \label{fig:rot_zams}
\end{figure*}

\begin{figure*}
    \centering
    \begin{tabular}{lll}
     \includegraphics[width=0.32\textwidth, page=1, trim=0.65cm 0.65cm 0.65cm 0.65cm, clip]{plots/Residual_rot_Xc_700.pdf}
    \includegraphics[width=0.32\textwidth, page=2, trim=0.65cm 0.65cm 0.65cm 0.65cm, clip]{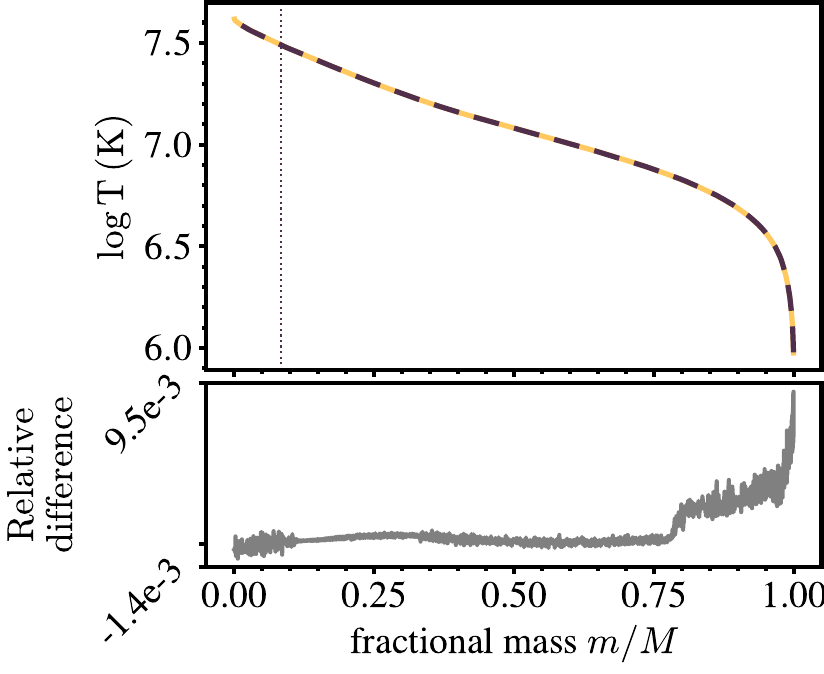}
    \includegraphics[width=0.32\textwidth, page=3, trim=0.65cm 0.65cm 0.65cm 0.65cm, clip]{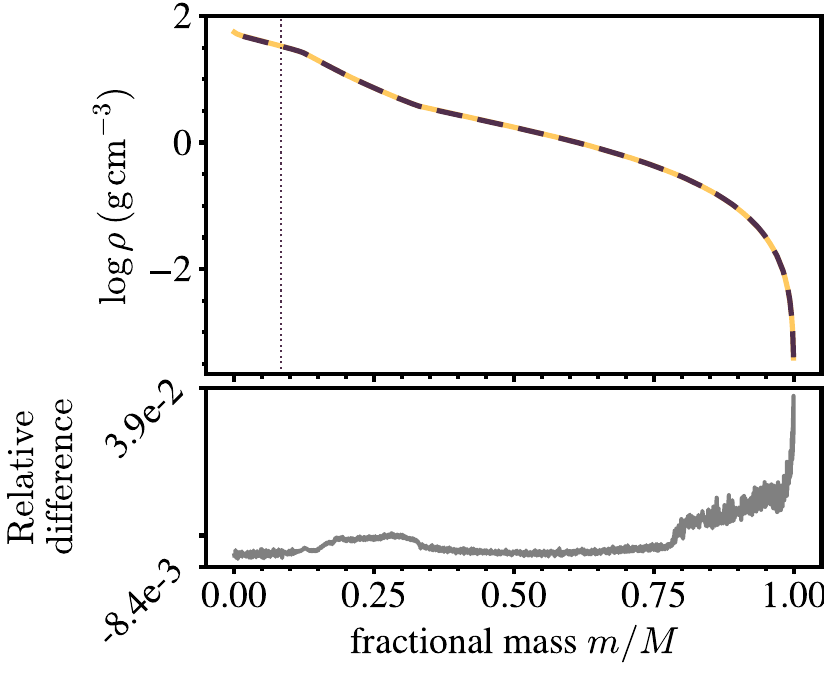}
    \\
    \includegraphics[width=0.32\textwidth, page=8, trim=0.65cm 0.65cm 0.65cm 0.65cm, clip]{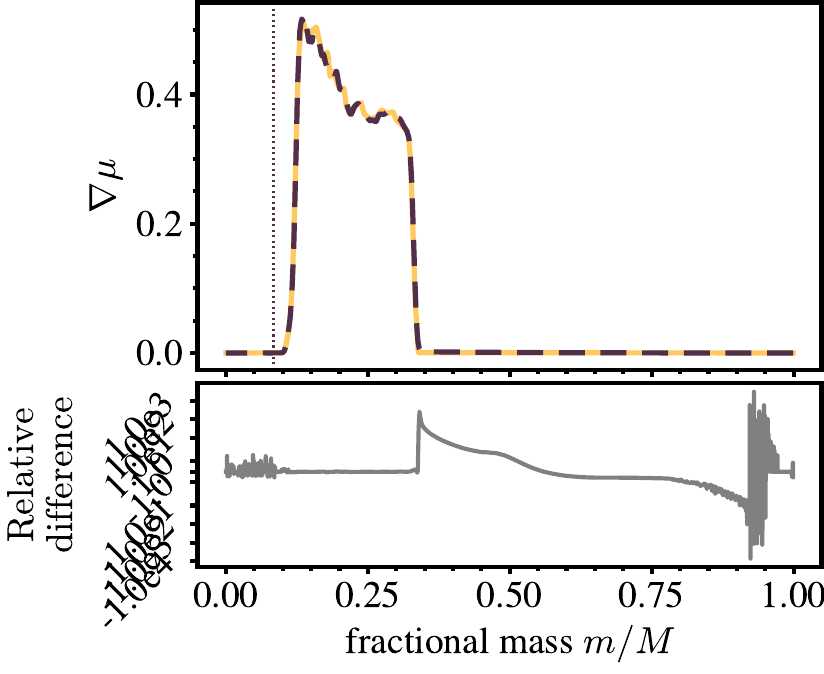}
    \includegraphics[width=0.32\textwidth, page=9, trim=0.65cm 0.65cm 0.65cm 0.65cm, clip]{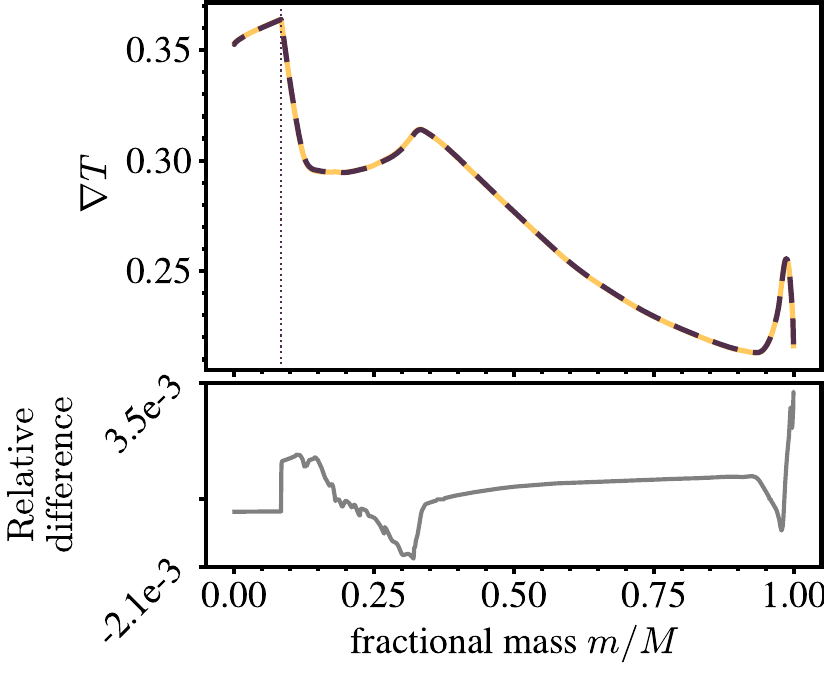}
    \includegraphics[width=0.32\textwidth, page=10, trim=0.65cm 0.65cm 0.65cm 0.65cm, clip]{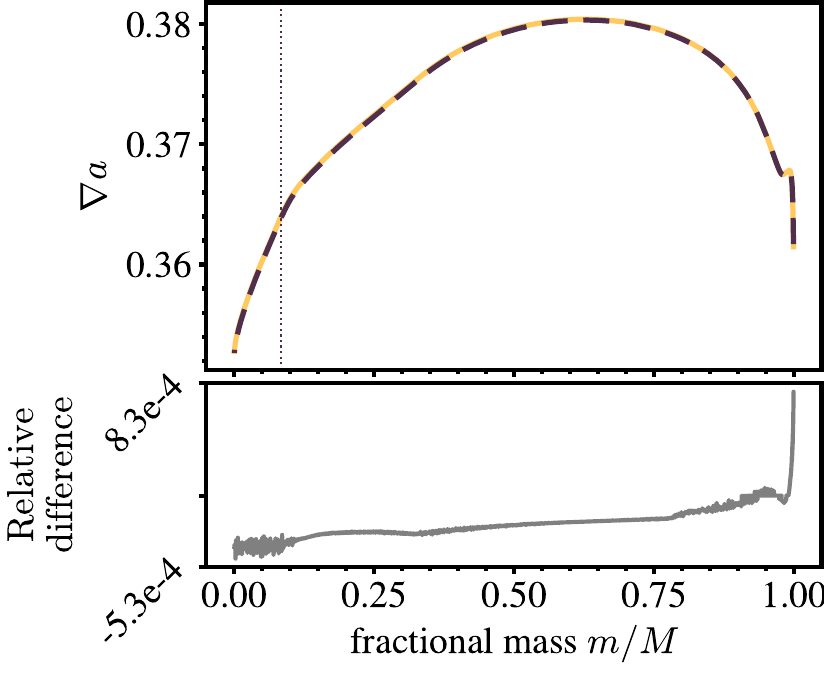}  
    \\
     \includegraphics[width=0.32\textwidth, page=4, trim=0.65cm 0.65cm 0.65cm 0.65cm, clip]{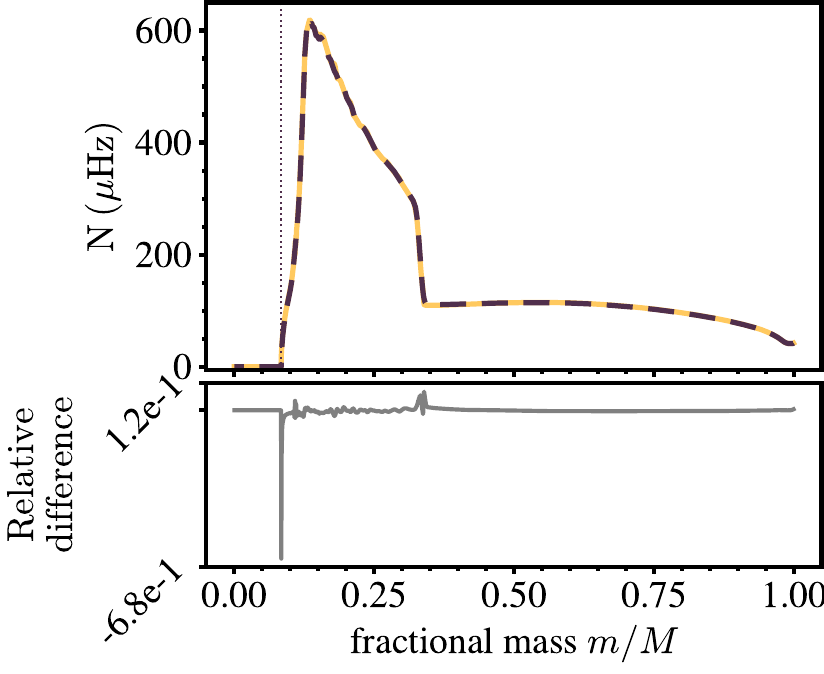}  
    \includegraphics[width=0.32\textwidth, page=5, trim=0.65cm 0.65cm 0.65cm 0.65cm, clip]{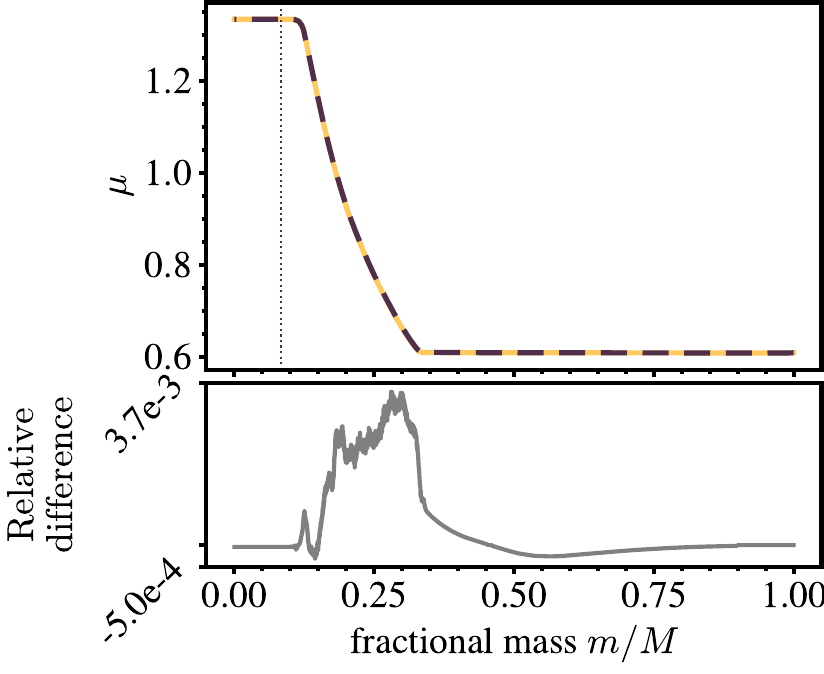}
    
    \end{tabular}
   
    \caption{Same as Fig.~\ref{fig:rot_zams} but for TAMS. Unlike rotating models with R-MLT in the convective regions (Fig.~\ref{fig:rotmlt_tams}), evolution during the main sequence causes hardly any increase in the differences between the parameters for rotating and non-rotating sets of models.}
    
    \label{fig:rot_tams}
\end{figure*}

\begin{figure}
\centering
    \includegraphics[width=0.45\textwidth]{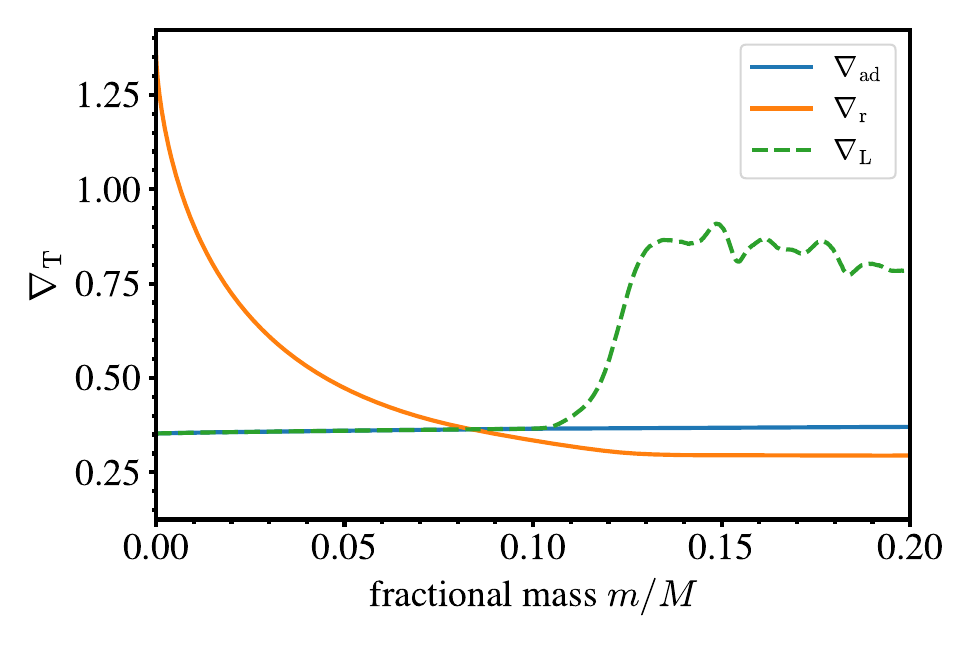} 
    \caption{{Temperature gradient in the interior of the 5\Msun{} model at the TAMS with initial rotation frequency 20 percent of the critical frequency at the surface. The radiative temperature gradient remains smaller than the adiabatic gradient throughout the radiative interior, except within the fully convective core region identified by the Ledoux criterion. Consequently, semiconvection does not develop.}}
    \label{fig:semiconvection}
\end{figure}

\end{appendix}

\end{document}